\documentclass[preprint,showpacs,preprintnumbers]{revtex4}
\usepackage{graphicx}
\usepackage{dcolumn}
\usepackage{bm}
\usepackage{subfigure}

\newcommand{\be}{\begin{equation}}
\newcommand{\en}{\end{equation}}
\newcommand{\bea}{\begin{eqnarray}}
\newcommand{\ena}{\end{eqnarray}}
\begin{document}

\title{ Intermediate inflation on the brane and warped DGP models}
\author{Ram\'on Herrera}
\email{ramon.herrera@ucv.cl}
\affiliation{Instituto de F\'{\i}sica, Pontificia Universidad Cat\'{o}lica de Valpara%
\'{\i}so, Casilla 4059, Valpara\'{\i}so, Chile.}
\author{Marco Olivares}
\email{marco.olivares@ucv.cl}
\affiliation{Instituto de F\'{\i}sica, Pontificia Universidad Cat\'{o}lica de Valpara%
\'{\i}so, Casilla 4059, Valpara\'{\i}so, Chile.}
\date{\today }
\author{Nelson Videla}
\email{nelson.videla@ucv.cl}
\affiliation{Instituto de F\'{\i}sica, Pontificia Universidad Cat\'{o}lica de Valpara%
\'{\i}so, Casilla 4059, Valpara\'{\i}so, Chile.}

\begin{abstract}
Brane and warped Dvali-Gabadadze-Porrati  inflationary universe
models in the context of intermediate inflation are studied. In both
models we consider that the energy density is a standard scalar
field and we discuss the evolution of the universe during the
slow-roll inflation. Also, we describe the conditions for these
models. The corresponding Wilkinson Microwave Anisotropy Probe seven
year data are utilized to fix some parameters in our models. Our
results are compared to those found in General Relativity.
\end{abstract}

\maketitle



\section{Introduction}

It is well known that the inflationary model is to date the most
compelling solution to many long-standing problems of the standard
Big Bang model (horizon, flatness, monopoles,
etc.)\cite{guth,infla}. The most significant feature of the
inflationary universe model is that it supplies a causal
interpretation of the origin of the observed anisotropy of the
cosmic microwave background radiation (CMB) and the distribution
of large scale structure in the universe\cite{astro,astro2}.

As regards exact solutions, the acceleration of the universe due
to an exponential potential is often called power-law inflation,
since the scale factor has an evolution  power law type
\cite{power}. In addition, de-Sitter inflationary universe is
produced by a constant scalar potential, see Ref.\cite{guth}.
Following Ref.\cite{Barrow1}, exact solutions can also be found
for the scenario of intermediate inflation. In this inflationary
model the scale factor, $a(t)$, increments as
\begin{equation}
a=\exp[\,A\,t^{f}],  \label{at}
\end{equation}
where $A$ and $f$ are two constants; $A>0$ and $0<f<1$
\cite{Barrow1}. The expansion of this scale factor  is slower than
de-Sitter inflation, but faster than power law inflation, this is
the reason why it is known as "intermediate". It is well known
that the intermediate inflationary model was originally formulated
as an exact solution, but it may be best developed  from the
slow-roll approximation.  Considering  this approximation, it is
possible to obtain a spectral index $n_s\sim 1$. In particular,
the value $n_s=1$ (Harrizon-Zel'dovich spectrum) is found for the
value of the parameter  $f=2/3$ \cite{Barrow2}, but this value of
the spectral index is disfavored by the current Wilkinson
Microwave Anisotropy Probe (WMAP) observational
data\cite{astro,astro2}. Also, an important observational quantity
in the intermediate model, is the tensor to scalar ratio $r$,
which is significantly $r\neq 0$\cite{ratior,Barrow3} for values
of $f\neq 2/3$. Other motivation to consider this type of the
expansion comes from string/M-theory, which appears to be relevant
in the low-energy string effective action{\cite{KM,ART}} (see
also, Refs.\cite{BD,Varios1,Varios2,Sanyal}). These theories can
be utilized to solve the initial singularity and describe the
present acceleration in the universe, among others\cite{new}. In
this way, the intermediate inflation model may be derived from an
effective theory at low dimensions of a fundamental string theory.

On the other hand, implications of string/M-theory to
Friedmann-Robertson-Walker (FRW) cosmological models have
currently attracted a great deal of attention; specifically some
were concerned with brane–antibrane configurations such as
space-like branes and the implications to cosmology\cite{sen1}. In
this configuration the standard model of particles is confined to
the brane, while gravitation propagates into the bulk space-time.
Here, the effect of extra dimensions induces extra terms in the
Friedmann equation \cite{1,3,8}. In particular, the cosmological
Randall-Sundrum (RS) type II scenario has received great attention
in the last years\cite{RS}.
 This alternative to Einstein's general relativity
cosmological models is called brane-world cosmology. For a
comprehensible  review of brane-cosmology, see e.g.
Refs.\cite{4,5,M}.

In the Dvali-Gabadadze-Porrati (DGP) model \cite{DGP} the induced
gravity brane world was presented as a replacement  to the RS
one-brane model \cite{RS}. Here,  general relativity was
recuperated, also despite an infinite extra dimension, but without
warping in 5-dimensional Minkowski space-time. In this model, the
gravitational actions  on the brane are controlled by the rival
between the 5D curvature scalar in the bulk and the 4D curvature
scalar on the brane. In comparison  to the RS case with high
energy modifications to general relativity, the DGP model produced
a low energy modification. In the DGP brane, in relation to the
embedding of the brane in the bulk, there are present two branches
of background solutions, because there are two different ways to
embed the 4-dimensional brane into the 5-dimensional space-time.
For a comprehensible review of the phenomenology of DGP model,
see\cite{Lue}; and inflation models and reheating in this scenario
were studied in Refs.\cite{Bou,Pap, Rong, Rong2,mas}.

The target of this work is to present two models of inflationary
universes  in the context of intermediate inflation, namely, (i)
when the scalar field is confined to the brane and (ii) when the
scalar field is confined on a  warped DGP.
We shall resort to the seven-year data of the Wilkinson Microwave
Anisotropy Probe (WMAP) to restrict the coefficients in both
models. In particular, we find constraints on the fundamental
parameters.

The outline of the paper is as follows. The next section presents
the
brane-intermediate Inflationary scenario for our model and in  subsection \ref%
{sectpert} deals with the calculations of cosmological
perturbations. In Section \ref{DGP}, we study the warped DGP model
in the context of the intermediate inflation and in subsection
\ref{PDGP} we also determine the corresponding cosmological
perturbation. Finally, in Sect.\ref{conclu} we summarize our
finding. We chose units so that $c=\hbar =1$.

\section{ Intermediate inflation on the brane}

\subsection{The model and the basic equations}
We consider the five-dimensional brane scenario, in which the flat
Friedmann equation is given by\cite{2,3}
\begin{equation}
H^2=\kappa\,\rho\left[1+\frac{\rho}{2\tau}\right]+\frac{\Lambda_4}{3}+\frac{\xi}{a^4},
\label{eq1}\end{equation} where  $H=\dot{a}/a$ denotes the Hubble
parameter, $a$ represents the scale factor and  $\rho$ is the
matter field confined to the brane. $\Lambda_4$ is the
4-dimensional cosmological constant and  the constant $\kappa=8\pi
G/3=8\pi/3m_p^2$.  Here, the term  $\xi/a^4$ denotes the influence
of the bulk gravitons on the brane, where  $\xi$ is an integration
constant. The parameter $\tau$ is the brane tension and is relates
the 4- and 5-dimensional Planck masses through
$m_p=\sqrt{3M_5^6/(4\pi\tau)}$. From nucleosynthesis, the brane
tension  is constrained by  the value $\tau
>$ (1MeV)$^4$ \cite{Cline}.  Nevertheless, a stronger constraint for the brane
tension results  from current tests for deviation from Newton`s
law, in which $\tau\geq $(10
 TeV)$^4$\cite{test1,test2}.

In the following, we will assume  that the
 cosmological constant $\Lambda_4=0$, and once
inflation begins, the  quantity  $\xi/a^4$ will rapidly become
unimportant, so that the Friedmann Eq.(\ref{eq1}) results in

\begin{equation}
H^{2}=\kappa\,\rho\left[1+\frac{\rho}{2\tau}\right].  \label{HC}
\end{equation}


In the following, we  consider  that  the matter field $\rho$ is a
standard scalar field $\phi $, in which the energy density $\rho=\frac{\dot{\phi}^{2}}{2}+V(\phi )$, and the pressure  $P=%
\frac{\dot{\phi}^{2}}{2}-V(\phi )$, where $V(\phi )=V$ is the
scalar potential.  We assume that the scalar field $\phi$ is
confined to the brane, so that its field equation has  the
standard form
\begin{equation}
\dot{\rho}+3H(\rho +P)=0,  \label{key_01}
\end{equation}%
or equivalently
\begin{equation}
\ddot{\phi}+3H\dot{\phi}+V'=0,  \label{ecdf}
\end{equation}%
where $V'=\partial V(\phi)/\partial \phi$ and  the dots mean
derivatives with respect to the cosmological time.

In the following, we will not consider  Eq.(\ref{HC}) in the lower
energy limit ($\rho\ll\tau$) or in the high energy limit
($\rho\gg\tau$) as our initiating point, instead we will study the
full Eq.(\ref{HC}).


Combining  Eqs.(\ref{HC}) and (\ref{key_01}) the square of the velocity $\dot{\phi}^2$ of the scalar field becomes%
\begin{equation}
\dot{\phi}^{2}=\frac{2}{3\kappa}\frac{(-\dot{H})}{\sqrt{1+\frac{2
H^{2}}{\kappa\tau }}},  \label{phipoint}
\end{equation}%
and the effective potential can be written
\begin{equation}
V=\tau \left( -1+\sqrt{1+\frac{2H^{2}}{\kappa \tau }}\right) -\frac{1}{3}%
\frac{\left( -\dot{H}\right) }{\kappa \sqrt{1+\frac{2H^{2}}{\kappa \tau }}}.
\label{efectivepot}
\end{equation}%
Note that in the low-energy limit, $\rho\ll \tau$,
 the expressions for $\dot{\phi}^{2}$ and the scalar
potential $V$ given by Eqs.(\ref{phipoint}) and
(\ref{efectivepot}),
reduce  to the standard inflation, where $%
\dot{\phi}^{2}=2(-\dot{H})/3\kappa $ and
$V=(3H^{2}+\dot{H})/3\kappa $, see Ref.\cite{Barrow3}.  Note also
that in the high energy limit, $\rho\gg \tau$,  the values of
$\dot{\phi}^{2}$ and the
scalar potential $V$ reduce to the brane world cosmology in which, $%
\dot{\phi}^{2}=\sqrt{2\tau/\kappa}\,\frac{(-\dot{H})}{3 H} $ and
$V=\sqrt{2\tau/\kappa}\,[H-\frac{\dot{H}}{6 H}]$\cite{nelson}.

The solution for the standard scalar field $\phi $ at late times
can be found from Eqs.(\ref{at}) and (\ref{phipoint})  as
\begin{equation}
\phi (t)-\phi _{0}=\frac{\mathcal{B}[t]}{K},  \label{fisol}
\end{equation}%
where $\phi(t=0)=\phi_0$ is an integration constant, $K$ is a
constant equal to
$$
K= \left[ \sqrt{\frac{3}{1-f}}\left( \tau ^{f}\left( \frac{\kappa }{2}%
\right) ^{2-f}\right) ^{\frac{1}{4}}\frac{1}{(Af)^{2}}\right] ^{\frac{1}{1-f}}, %
$$
and the function $\mathcal{B}[t]$, represents the incomplete Beta
function \cite{Libro},  given by
$$
\mathcal{B}[t]= B\left[\frac{\kappa \tau\,t^{2(1-f)} }{2(Af)^{2}};\frac{f}{%
4(f-1)},\frac{3}{4}\right].
$$

From Eq.(\ref{fisol}), the Hubble parameter as a function of the
scalar field, $\phi $, becomes $ H(\phi
)=Af(\mathcal{B}^{-1}[K(\,\phi)])^{f-1}$, where $\mathcal{B}^{-1}$
represents the inverse function of the incomplete Beta function
and without loss of generality we have used $\phi_0=0$.

Considering  the slow-roll approximation, only the first term of
the effective potential given by Eq.(\ref{efectivepot})
predominates at large values  of $\phi $. In this way, combining  Eqs.(%
\ref{efectivepot}) and (\ref{fisol}), the effective potential as a
function of the scalar field, $\phi $, results in

\begin{equation}
V(\phi )\simeq \tau \left( -1+\sqrt{1+\frac{2(Af)^{2}(\mathcal{B}%
^{-1}[K(\,\phi )])^{-2(1-f)}}{\kappa \tau }}\right) .
\label{potfi}
\end{equation}%
Here, we noted that we would have obtained the same effective potential represented  by Eq.(%
\ref{potfi}), considering the  slow-roll conditions, in which $\dot{\phi}%
^{2}\ll V(\phi )$ and $\ddot{\phi}\ll 3H\dot{\phi}$. Also, we
noted that in low-energy limit $V(\phi)\propto
\,\phi^{-4(f^{-1}-1)}$ \cite{Barrow3} and in the high energy limit
$V(\phi)\propto\,\phi^{2(f-1)/3}$, see Ref.\cite{nelson}.

The dimensionless slow-roll parameters  can be expressed as $
\varepsilon \equiv -\frac{\dot{H}}{H^{2}}=\frac{1-f}{Af(\mathcal{B}%
^{-1}[K(\,\phi )])^{f}} $ and $
\eta \equiv -\frac{\ddot{H}}{H\dot{H}}=\frac{2-f}{Af(\mathcal{B}%
^{-1}[K(\,\phi )])^{f}}$, respectively.  The inflationary phase
takes place whenever  $\varepsilon<1$ (or analogously
$\ddot{a}>0$). Consequently, the
condition for inflation to occur is satisfied when  $\phi >%
\frac{1}{K}\,\mathcal{B}\left[ \left( \frac{1-f}{Af}\right)
^{1/f}\right] $. Also, considering that inflation begins at the
earliest possible phase (when $\varepsilon =1$) see
Ref.\cite{Barrow3}, then
$\phi _{1}$, can be expressed as $\phi _{1}=\frac{1}{K}\,\mathcal{B}%
\left[ \left( \frac{1-f}{Af}\right) ^{1/f}\right] .$

Finally,  the number of e-folds $N$ between two values of
cosmological times $t_{1}$ and $t_{2}$ or analogously between two
different values $\phi _{1}$ and $\phi _{2}$  becomes
\begin{equation}
N=\int_{t_{1}}^{t_{2}}\,H\,dt=A\,\left[ (t_{2})^{f}-(t_{1})^{f}\right] =A\,%
\left[ (\mathcal{B}^{-1}[K(\,\phi _{2})])^{f}-(\mathcal{B}%
^{-1}[K(\,\phi _{1})])^{f}\right] .  \label{N}
\end{equation}

Here, we have used Eq.(\ref{fisol}).

\subsection{Cosmological perturbations\label{sectpert}}

In this subsection we will describe scalar and tensor
perturbations, for our model. For a standard scalar field $\phi$
the power spectrum ${\mathcal{P}_{\mathcal{R}}}$ of the curvature
perturbations is given in the slow-roll approximation by the following expression %
 \cite{4}:

\begin{equation}
{\mathcal{P}_{\mathcal{R}}}\simeq \left( \frac{H^{2}}{2\pi
\dot{\phi}}\right) ^{2} =
\frac{3\kappa }{8\pi ^{2}}H^{4}\left( -%
\dot{H}\right) ^{-1}\sqrt{1+\frac{2H^{2}}{\kappa \tau }}.
\label{Ph}
\end{equation}%
Here, we have used Eq.(\ref{phipoint}). The power spectrum can be
written   equivalently in terms of the standard scalar field $\phi
$ as
\begin{equation}
{\mathcal{P}_{\mathcal{R}}}= \frac{3\kappa }{8\pi ^{2}}\frac{(Af)^{3}}{%
1-f}(\mathcal{B}^{-1}[K(\,\phi )])^{-(2-3f)}\sqrt{1+\frac{%
2(Af)^{2}(\mathcal{B}^{-1}[K(\,\phi )])^{-2(1-f)}}{\kappa \tau }}.
\label{Pfi}
\end{equation}

Combining Eqs.(\ref{N}) and (\ref{Pfi}),
${\mathcal{P}_{\mathcal{R}}}$ can be expressed in terms of the
number of e-folds $N$, to give
\begin{equation}
{\mathcal{P}_{\mathcal{R}}}= \frac{3\kappa }{8\pi ^{2}}\frac{(Af)^{3}}{%
1-f}\left[ \frac{Af}{1+f(N-1)}\right] ^{\frac{2-3f}{f}}\sqrt{1+\frac{%
2(Af)^{2}\left[ \frac{Af}{1+f(N-1)}\right] ^{\frac{2(1-f)}{f}}}{\kappa \tau }%
}.  \label{PN}
\end{equation}

Numerically from Eq.(\ref{PN}), we obtained  a constraint for the
parameter $A$. In fact, we can obtain the value of the parameter
$A$ for  given values of $f$ and the brane tension $\tau$ when
number $N$ and the  power spectrum ${\mathcal{P}_{\mathcal{R}}}$
are given. In particular, for the values
${\mathcal{P}_{\mathcal{R}}}=2.4\times 10^{-9}$, $N=60$, $f=1/2$
and $m_p=1$, we obtained  for the brane tension: $\tau=10^{-6}$,
which corresponds to the parameter $A\simeq 3.694\times 10^{-2}$,
 $\tau=10^{-8}$, which corresponds to $A\simeq 3.692\times
10^{-2}$, and  $\tau=10^{-10}$, which corresponds to $A\simeq
3.587\times 10^{-2}$.

The scalar spectral index $n_{s}$ is given by $n_{s}-1=\frac{d\ln \,{%
\mathcal{P}_{R}}}{d\ln k}$, where the interval in wave number is related to
the number of e-folds $N$ by the relation $d\ln k(\phi )=-dN(\phi )$. From
Eq.(\ref{Pfi}) the scalar spectral index can be written as
\begin{equation}
n_{s}= 1-\frac{(2-3f)}{Af}(\mathcal{B}^{-1}[K(\,\phi )])^{-f}-%
\frac{2Af(1-f)}{\kappa\tau }\frac{(\mathcal{B}^{-1}[K(\,\phi
)])^{-(2-f)}}{\left(1 +\frac{2(Af)^{2}(\mathcal{B}^{-1}[K(\,\phi
)])^{-2(1-f)}}{\kappa\tau }\right)}.  \label{nsa}
\end{equation}
Here, we noted  from Eq.(\ref{nsa}) that $n_s\neq 1$, for $f=2/3$
(recall  that $1>f>0$). However, as occurs in the standard
intermediate model, $n_s=1$ for the value $f=2/3$, where the scale
factor increases as $a(t)\sim e^{t^{2/3}}$, see
Ref.\cite{Barrow3}.

On the other hand, the tensor-perturbation during inflation would
produce gravitational waves. This perturbation in cosmology is
more involved in our case, since in the brane-world gravitons
propagate in the bulk. The amplitude of the tensor perturbations
was calculated in Ref.\cite{t}, where
\begin{equation}
{\mathcal{P}}%
_{g}=24\kappa \,\left( \frac{H}{2\pi }\right)
^{2}F^{2}(x)\label{PGB}.
\end{equation}
In our model we get
\begin{equation}
{\mathcal{P}}_{g}=\frac{6\kappa (Af)^{2}}{\pi ^{2}}(\mathcal{B}%
^{-1}[K(\,\phi )])^{-2(1-f)}F^{2}(x),  \label{ag}
\end{equation}%
where $x=Hm_{p}\sqrt{3/(4\pi \tau )}=(\mathcal{B}^{-1}[K(\,\phi
)])^{-(1-f)}\sqrt{\frac{3(Afm_{p})^{2}}{4\pi \tau }}$ and the
function $F(x)$ is given by
\begin{equation}
F(x)=\left[ \sqrt{1+x^{2}}-x^{2}\sinh ^{-1}(1/x)\right] ^{-1/2},
\label{Fx}
\end{equation}
here, the function $F(x)$, appeared from the normalization of a zero-mode%
\cite{t}.

On the other hand, an important observational quantity is the
tensor to scalar ratio $r$, which is defined as
$r=\left(\frac{{\mathcal{P}}_g}{P_{\mathcal{R}}}\right)$.
Combining Eqs.(\ref{Pfi}) and (\ref{ag}), the tensor-scalar ratio,
$r$, is given by
\begin{equation}
r=\left( \frac{{\mathcal{P}}_{g}}{P_{\mathcal{R}}}\right) =\frac{16(1-f)}{Af%
}\frac{(\mathcal{B}^{-1}[K(\,\phi )])^{-f}}{\sqrt{1+\frac{%
2(Af)^{2}(\mathcal{B}^{-1}[K(\,\phi )])^{-2(1-f)}}{\kappa \tau }}}%
\,F^{2}(\phi),  \label{Rk1}
\end{equation}%
also, the relation between the tensor-scalar ratio $r$ and the
number e-folds $N$ can be written as

\begin{equation}
r(N)=\frac{16(1-f)}{Af}\frac{\left[ \frac{Af}{1+f(N-1)}\right] }{\sqrt{1+\frac{%
2(Af)^{2\left[ \frac{Af}{1+f(N-1)}\right] \frac{2(1-f)}{f}}}{\kappa \tau }}}%
\,F^{2}(N).  \label{RN}
\end{equation}

\begin{figure}[th]
\includegraphics[width=3.3in,angle=0,clip=true]{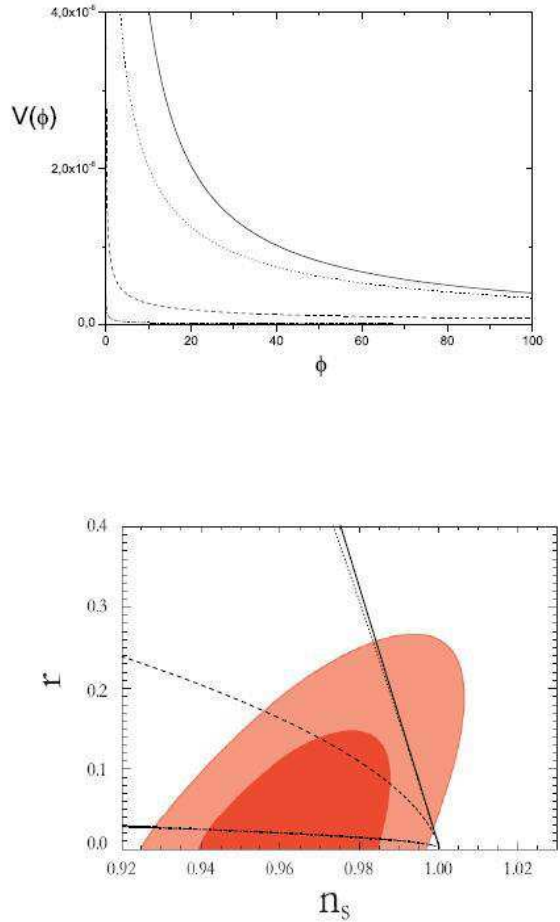}
\caption{ The upper panel shows the evolution of the scalar
potential $V$ versus the scalar field $\phi$. The lower panel
shows the contour plot for the parameter $r$  as a function of the
$n_s$ . In both panels we considered  different values of the
brane tension $\tau$.
From left to right, the dot-dashed, dashed, dotted and solid lines
are for the brane tension $\tau=10^{-10}$, $\tau=10^{-8}$,
$\tau=10^{-6}$, and the standard intermediate model
($\tau\rightarrow\infty$), respectively. In both panels we have
taken the values  $f=1/2$, $m_p=1$, and $A\simeq 3.587\times
10^{-2}; 3.692\times 10^{-2}; 3.694\times 10^{-2}$, respectively.
\label{1}}
\end{figure}

In Fig.(\ref{1}), the upper panel shows the evolution of the
effective potential $V(\phi)$ versus the scalar field $\phi$, and
the lower panel shows the contour plot for the tensor-scalar ratio
$r$ as a function of the $n_s$. In both panels we studied
different values of the brane tension $\tau$. In particular, the
dot-dashed, dashed, dotted, and solid lines are for the brane
tension; $\tau=10^{-10}$, $\tau=10^{-8}$, $\tau=10^{-6}$, and the
standard intermediate model ($\tau\rightarrow\infty$),
respectively. From the upper panel, we noted that, when we
increased the brane tension $\tau$, the effective potential graphs
present a small displacement with respect to the scalar field
$\phi$, when compared to the results of the potential
 obtained in the standard
intermediate model in which  $V(\phi)\propto\phi^{-4(f^{-1}-1)}$.

In the lower panel of  Fig.(\ref{1}), we show the dependence of
the tensor-scalar ratio $r$ on the spectral index $n_s$. Following
Ref.\cite{astro}, we have 2-dimensional marginalized constraints
corresponding to 68$\%$ and 95$\%$ confidence levels on
inflationary parameters $r$ and $n_s$, the spectral index of
fluctuations, defined at $k_0$ = 0.002 Mpc$^{-1}$. In order to
write down values that relate the ratio $r$ and the index $n_s$,
we numerically solved Eqs. (\ref{nsa}) and (\ref{Rk1}). Also, we
have taken the values $f=1/2$, $m_p=1$ and for the parameter $A$
the values $A\simeq 3.587\times 10^{-2}; 3.692\times 10^{-2};
3.694\times 10^{-2}$, respectively. We noted that the values of
the brane tension $\tau=10^{-10}$ and $\tau=10^{-8}$ enter the
marginalized constraints corresponding to 68$\%$ confidence
levels. However, the value $\tau=10^{-6}$ only enters on 95$\%$
confidence level, as could be seen from the lower panel of
Fig.(\ref{1}). Also, for the value of the  brane tension
$\tau=10^{-6}$, we numerically observed that the consistency
relations, $n_s = n_s(r)$, present a small displacement in
relation to the standard intermediate model. We  noted that when
we increase the value of the parameter $\tau$, for values of
$\tau>10^{-6}$, we observed that the contour plot in the $r-n_s$
plane becomes similar to the standard intermediate model. Also, we
observed that the incorporation of the brane tension parameter
gives us a freedom that allows us to modify the standard
intermediate model by simply modifying the corresponding value of
the parameter $\tau$, e.g., on the contour plot in the $r-n_s$
plane. In this form, our model is less restricted than the
standard intermediate-model.

\section{Intermediate inflation on the warped DGP brane}\label{DGP}

\subsection{The model and the basic equations}

We start by writing down  the  Friedmann equation on the warped
DGP brane \cite{Ke}
\begin{equation}
 \label{H1}
 H^2+\frac{k}{a^2}={1\over 3\mu^2}\Bigl[\,\rho+\rho_0\bigl(1 +
\epsilon{\cal A}(\rho, a)\bigr)\,\Bigr],
\end{equation}
where $k$ is the constant curvature of the three-space of the  FRW
metric and
 the $\mu$ parameter   denotes the strength of the induced gravity term on the
 brane, with the parameter $\mu=0$ giving the RS model. The  $\epsilon$ parameter becomes  either $+1$ or $-1$,
 corresponding  to the two branches of this model. We will denote
 $\epsilon=+1$as the positive branch and  the value $\epsilon=-1$
 as the negative branch.
The function ${\cal A}(\rho,a)$ is given  by

$$
 {\cal A}(\rho,a)=\left[{\cal A}_0^2+{2\eta\over
\rho_0}\left(\rho-\mu^2 {{\cal E}_0\over a^4}
\right)\right]^{1\over 2},
 $$
where the constants ${\cal A}_0$, $\rho_0$, and $\eta$, are
defined as
 \begin{equation}
 {\cal A}_0=\sqrt{1-2\eta{\mu^2\Lambda\over \rho_0}},\,\;\;\rho_0=\tau+6{m_5^6\over
 \mu^2}\;\;
\,\,\,\mbox{and}\,\,\,\,\,\,\eta={6m_5^6\over \rho_0\mu^2}
~~~(0<\eta\leq 1).\label{Ao}
 \end{equation}
Here, the constant $\Lambda$ is the effective 4-dimensional
cosmological constant on the brane and
 is defined by
 $
\Lambda={1\over 2} ({}^{(5)}\Lambda+{1\over 6}\kappa_5^4\,\tau^2),
 $
 in which $\kappa_5^2=m_5^{-3}$ is the 5-dimensional gravitational  constant, ${}^{(5)}\Lambda$ is the 5-dimensional cosmological constant
 in the bulk, as before
 $\tau$ is the brane tension and ${\cal E}_0$ is a constant related to Weyl
 radiation.

In the following,  we will neglect the curvature term and the dark
radiation term during inflation and  the function ${\cal
A}(\rho,a)\approx{\cal A}(\rho)\approx\left[{\cal
A}_0^2+{2\eta\rho\over \rho_0}\right]^{1\over 2}$. In this way,
the Friedmann Eq.(\ref{H1}) takes the form
\begin{equation}
 \label{H2}
 H^2={1\over 3\mu^2}\left[\rho+\rho_0+\epsilon\left({\cal
 A}_0^2+\frac{2\eta\rho}{\rho_0}\right)^{1/2}\right].
\end{equation}
In particular, for the DGP scenario, one has ${\cal
 A}_0=1$ and the parameter $\eta=1$.

 Following Ref.\cite{Ke}, we noted  that in the ultra high energy limit where
 $ \rho \gg \rho_0 \gg \tau$ and ${\cal A}_0=1$, the effective Friedmann Eq.(\ref{H2})
 becomes
  $
 H^2 = \frac{1}{3\mu^2}\left( \rho +\epsilon \sqrt{2\rho
 \rho_0}\right).
 $
 Also, in the
 intermediate energy region in which $\rho \ll \rho_0 $ but $\rho \gg
 \tau$, for  $\epsilon =-1$, the effective Friedmann
 equation can be rewritten by
 $
 H^2  = \frac{\tau}{18m_5^6}\left( \rho
+\frac{\rho^2}{2\tau} -\frac{\mu^2 \tau}{6m_5^6}\rho
-\frac{\mu^2}{4m_5^6}\rho^2\right).
 $
 Finally,  in low energy limit in which  $\rho \ll \tau \ll \rho_0$
 the Friedmann
 Eq.(\ref{H2}) becomes
 $
 H^2 = \frac{1}{3\mu^2_p}\left [\rho + {\cal O}\left
(\frac{\rho}{\rho_0}\right )^2\right],
$
 where the effective 4-dimensional Planck mass is defined by $\mu^2_p= \mu^2/(1-\eta)$.

Analogously, as  before we will consider that the energy density
$\rho$ is a standard scalar field $\phi$ satisfying the continuity
equation given by Eq.(\ref{key_01}). In this form, the square of
the velocity $\dot{\phi}^2$ of the scalar field, considering
Eqs.(\ref{key_01}) and (\ref{H2}), becomes

\begin{equation}
 \dot{\phi}^2= 2\mu^2(-\dot{H})\left[1-
\epsilon\left(\alpha+\beta H^2\right)^{-1/2} \right],\label{a4}
\end{equation}
where the constants $\alpha$ and $\beta$ are defined by
$$
\alpha= 1+\frac{{\cal{A}}_0 ^{2}}{\eta^2}-\frac{2}{\eta}
,\;\;\;\mbox{and} \,\,\; \beta=\frac{6\mu^2}{\eta\rho_0}.
$$

 The effective potential as function of the $H$ and $\dot{H}$, can be obtained from
Eqs.(\ref{H2}) and (\ref{a4}), and as a result is found

$$
V=\frac{\eta\rho_0 }{2}\left(\alpha+\beta H^2\right) \left[1-
\epsilon\left(\alpha+\beta H^2\right)^{-1/2} \right]^{2}+
$$
\begin{equation}
+\mu^2\dot{H}\left[1- \epsilon\left(\alpha+\beta H^2\right)^{-1/2}
\right]-\frac{{\cal{A}}_0 ^{2}\rho_0}{2\eta} .\label{a5}
\end{equation}


 Combining Eqs.(\ref{at}) and (\ref{a4})
we found a relation between the scalar field and cosmological time
given by
\begin{equation}
\phi(t)-\phi_0=\frac{F[t]}{K},\label{a7}
\end{equation}
where $\phi(t=0)=\phi_0$ is an integration  constant
 and the constant $K$, is given by
$$
K= \nu \left(\frac{1-f}{2 \mu^2 Af} \right)^{1/2} (\beta A^2
f^2)^{-a_f/2}.
$$
Here, the function $F[t]$ is the incomplete Lauricella function,
see Ref.\cite{Libro2} and is defined as

$$
F[t]=\left(\alpha+\frac{\beta}{t^{2(1-f)}
}\right)^{\frac{-\nu}{2}}
F^{(3)}_{D}[\nu;1+\frac{\nu}{2},1+\frac{\nu}{2},\frac{-1}{2}
,\nu+1,\sqrt{\alpha},-\sqrt{\alpha},
\epsilon\left(\alpha+\frac{\beta}{t^{2(1-f)}
}\right)^{\frac{-1}{2}} ],
$$
where $ \nu=f\,[2(1-f)]^{-1}$. The Hubble parameter as a function
of the scalar field, $\phi$, results in $
H(\phi)=Af\,(F^{-1}[K\phi])^{f-1}$. Here, we have used $\phi_0=0$.

As before, considering the slow-roll approximation, the scalar
potential given by Eq.(\ref{a5}) reduces to

\begin{equation}
V(\phi)\simeq \frac{\eta\rho_0 }{2}\left(\alpha+ \frac{\beta
A^2f^2}{(F^{-1}[K\phi])^{2(1-f)}}\right) \left[1-
\epsilon\left(\alpha+
 \frac{\beta A^2f^2}{(F^{-1}[K\phi])^{2(1-f)}}\right)^{-1/2}
\right]^{2} -\frac{{\cal{A}}_0 ^{2}\rho_0}{2\eta}.\label{a9}
\end{equation}

Again, introducing the dimensionless slow-roll parameters
$\varepsilon$ and $\eta$, we have $
\varepsilon=-\frac{\dot{H}}{H^2}=\frac{1-f}{Af(F^{-1}[K\phi])^{f}},
$ and $ \eta=-\frac{\ddot{H}}{H
\dot{H}}=\frac{2-f}{Af(F^{-1}[K\phi])^{f}}\,. $ As before, if we
consider that the inflationary scenario begins at the earliest
possible stage, in which $\varepsilon=1$, then the scalar field
$\phi_1$, is given by  $
\phi_{1}=\frac{1}{K}F\left[\left(\frac{1-f}{Af}\right)^{1/f}\right]$.
Also, the condition for inflation to occur is $\varepsilon<1$ or
equivalently   when
$\phi>\frac{1}{K}F\left[\left(\frac{1-f}{Af}\right)^{1/f}\right]$.

For the number of e-folds between two different values $\phi_1$
and $\phi_2$ of the scalar field we have
\begin{equation}
N=\int_{t_1}^{t_{2}}\,H\,dt=A\,\left[(F^{-1}[K\phi_{2}])^{f}-(F^{-1}[K\phi_{1}])^{f}\right].\label{N1}
\end{equation}
Here, we have considered  Eq.(\ref{a7}).


\subsection{Cosmological perturbations}\label{PDGP}

In the following, we will study  the power spectra of scalar and
tensor perturbations to the metric in our inflationary model. We
consider  the gauge invariant quantity $
\zeta=H\,+\frac{\delta\rho}{\dot{\rho}}$
\cite{Bardee,Malik:2008im}. Here, $\zeta$ defined on slices of
uniform density and
 reduces to the curvature perturbation. A fundamental
feature  sign of $\zeta$ is that it is nearly constant on
super-horizon scales. This property,  an effect of stress-energy
conservation,   does not depend on the gravitational
dynamics\cite{Bassett:2005xm} (see also, Ref.\cite{wand2}). In
this form, the power spectrum
related to the curvature spectrum, could be written as ${\mathcal{%
P}_{\mathcal{R}}}\simeq\langle\zeta^2\rangle$ and it stays
unchanged in the warped DGP model \cite{Bou,PPP}(see also
Ref.\cite{nunes}). Therefore, for the spatially flat gauge, we
have $\zeta=H\frac{\delta\phi}{\dot{\phi}}$, in which
$|\delta\phi|=H/2\pi$ \cite{infla}.

In this way, the power spectrum, considering  Eq.(\ref{a4}), is
given by

\begin{equation}
{\cal{P}_{\cal{R}}}\simeq\left(\frac{H^2}{2\pi\dot{\phi}}\right)^2=\frac{1}{8\pi^{2}\mu^2}H^{4}(-\dot{H})^{-1}\left[1-
\epsilon\left(\alpha+\beta H^2\right)^{-1/2} \right]^{-1},
\label{pd1}
\end{equation}
or equivalently in terms of the standard scalar field $\phi$
\begin{equation}
{\cal{P}_{\cal{R}}}=\frac{A^3f^3(1-f)^{-1}}{8\pi^{2}\mu^2(F^{-1}[K\phi])^{2-3f}}\left[1-
\epsilon\left(\alpha+ \frac{\beta
A^2f^2}{(F^{-1}[K\phi])^{2(1-f)}}\right)^{-1/2} \right]^{-1}.
\label{pd2}
\end{equation}

Analogously as before, the power spectrum,
${\mathcal{P}_{\mathcal{R}}}$, also can be expressed in terms of
the number of e-folds $N$, as

$$
{\cal{P}_{\cal{R}}}=\frac{A^3f^3}{8\pi^{2}\mu^2(1-f)}
\left[\frac{Af}{1+f(N-1)}\right]^{\frac{2-3f}{f}}\,\,\times
$$
\begin{equation}
\times\,\,\left[1-\epsilon\left(\alpha+\beta A^2f^2
\left[\frac{Af}{1+f(N-1)}\right]^{\frac{2(1-f)}{f}}\right)^{-1/2}
\right]^{-1} .\label{Pw}
\end{equation}

Again,  numerically from Eq.(\ref{Pw}), we found a constraint for
the parameter $A$. Analogously, as in the case of the RS model, we
can find the value of the parameter $A$ for given values of $f$,
$m_5/\mu$, $\epsilon$, and  $\eta$, when $N$ and
${\mathcal{P}_{\mathcal{R}}}$ are given. In the following, we will
assume  that  the effective 4-dimensional cosmological constant
$\Lambda=0$, which implies that is ${\cal A}_0=1$, see
Eq.(\ref{Ao}), and the parameter $\eta=0.99$, see Ref.\cite{PPP}.

In particular, for the values
${\mathcal{P}_{\mathcal{R}}}=2.4\times 10^{-9}$, $N=60$, $f=1/2$,
$\epsilon=-1$, $m_p=1$,  and the ratio $m_5/\mu=0.02$, see
Ref.\cite{PPP},   we found for  the parameter $A\simeq 2.69\times
10^{-2}$ and for the ratio $m_5/\mu=0.2$,  which corresponds to
$A\simeq 5.98\times 10^{-2}$.

On the other hand, the scalar spectral index $n_s$, considering
Eq.(\ref{pd2}), we get

\begin{equation}
n_s=
1-\frac{2-3f}{A\,f}(F^{-1}[K\phi])^{-f}+n_\epsilon,\label{nss1}
\end{equation}
where the correction, $n_\epsilon$, in the scalar spectral index
is given by
$$
n_\epsilon=\epsilon \frac{\beta Af(1-f)}{(F^{-1}[K\phi])^{2-f}}
\left(\alpha+ \frac{\beta
A^2f^2}{(F^{-1}[K\phi])^{2(1-f)}}\right)^{-3/2} \left[1-\epsilon
\left(\alpha+ \frac{\beta
A^2f^2}{(F^{-1}[K\phi])^{2(1-f)}}\right)^{-1/2} \right]^{-1}.
$$
We noted numerically from Eq.(\ref{nss1}) that the value of the
spectral index is $n_s\gg 1$ for the positive branch
$\epsilon=+1$, but this value of  $n_s$ is disfavored from the
observational data. In this form, intermediate inflation on a
warped DGP cannot exit in the positive branch $\epsilon=+1$. In
the following, we will consider the negative branch $\epsilon=-1$.

\begin{figure}[th]
\includegraphics[width=3.1in,angle=0,clip=true]{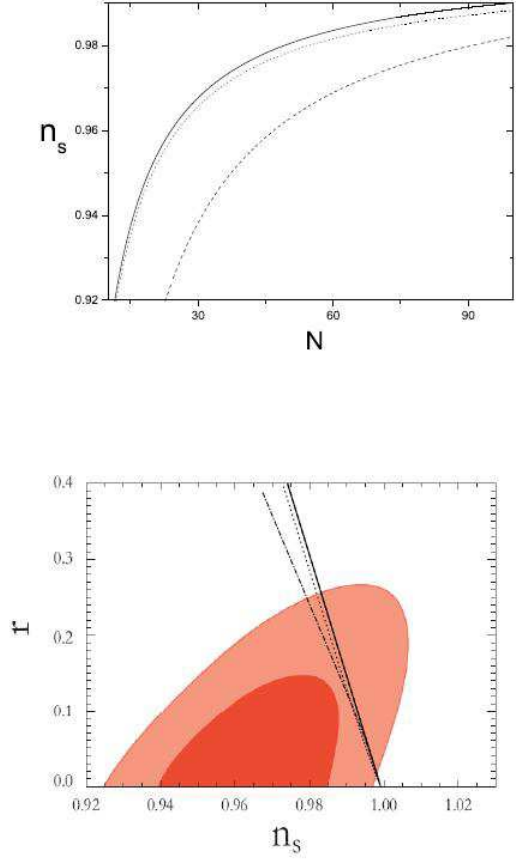}
\caption{  The upper panel shows the evolution of the scalar
spectrum index $n_s$ versus the number of e-folds $N$. The lower
panel shows the contour plot for the parameter $r$  as a function
of the $n_s$. Here, from WMAP seven-year data\cite{astro}, we have
2-dimensional marginalized constraints (68$\%$ and 95$\%$
confidence levels) on the inflationary parameters $r$ and $n_s$.
The dashed, dotted and solid  lines are for the  pairs
$(A=5.98\times10^{-2},m_5/\mu=0.2)$,
$(A=2.69\times10^{-2},m_5/\mu=0.02)$, and the standard
intermediate model \cite{Barrow3}, respectively. In both panels we
have taken the values $m_p=1$, $\eta=0.99$, $\epsilon=-1$,
${\cal{A}}_0=1$, and $f=1/2$. \label{2}}
\end{figure}

Combining Eqs.(\ref{N1}) and (\ref{nss1}), the scalar spectral
index can be re-expressed in terms of the number of e-folds as
\begin{equation}
n_s(N) =  1-\frac{2-3f}{1+f(N-1)}+n_\epsilon(N),\label{nsw}
\end{equation}
where the correction $n_\epsilon(N)$ becomes
$$
n_\epsilon(N)=\epsilon\beta Af(1-f)
\left[\frac{Af}{1+f(N-1)}\right]^{\frac{2-f}{f}}
\left(\alpha+\beta A^2f^2
\left[\frac{Af}{1+f(N-1)}\right]^{\frac{2(1-f)}{f}}\right)^{-3/2}\times
$$
$$
\left[1-\epsilon \left(\alpha+\beta A^2f^2
\left[\frac{Af}{1+f(N-1)}\right]^{\frac{2(1-f)}{f}}\right)^{-1/2}
\right]^{-1}.
$$

Following Ref.\cite{Bou} the amplitude of gravitational waves in
our model is given by
\begin{equation}
{\cal{P}}_g=\frac{64\pi}{m_p^2}\,\left(\frac{H}{2\pi}\right)^2\,G^2_{\gamma}(x),\label{Agg}
\end{equation}
where  the correction to standard 4D general relativity is
$G^{-2}_\gamma(x)=\gamma+(1-\gamma)F(x)^{-2}$ and the parameter
$\gamma=\mu^2/m_p^2$. Here, the function $F(x)$ is defined by
Eq.(\ref{Fx}) and $x=H/\bar{\mu}$, where $\bar{\mu}$ is the energy
scale associated with the bulk curvature\cite{Bou}. In particular,
for the brane cosmological constant to zero,
$\kappa_4^2=m_p^{-2}=\kappa_5^2\,\bar{\mu}\,(1-\gamma)$
\cite{Ke,Bou}. Note that the correction $G_\gamma^2$ coincides to
the expression the RS case when $\gamma\rightarrow 0$, see
Eq.(\ref{PGB}). Also, when $x\rightarrow 0$, we get
$G_\gamma^2\rightarrow 1$ and it reduces to the standard 4D
amplitude of gravitational waves in which
${\cal{P}}_g=\frac{64\pi}{m_p^2}\,\left(\frac{H}{2\pi}\right)^2$.

In this way, considering Eqs.(\ref{pd2}) and (\ref{Agg}) we may
write the tensor-scalar ratio as
$r=\left(\frac{{\cal{P}}_g}{P_{\cal R}}\right) $ and this ratio as
a function of the scalar field becomes
\begin{equation}
r
=\frac{128\pi\,(1-f)}{Af\,(F^{-1}[K\phi])^{f}}\,\frac{\mu^2}{m_p^2}
\left[1-\epsilon \left(\alpha+ \frac{\beta
A^2f^2}{(F^{-1}[K\phi])^{2(1-f)}}\right)^{-1/2}
\right]\,\,G_\gamma^2(\phi). \label{Rk}
\end{equation}

Analogously, as before we can write the tensor-scalar ratio as a
function of the number of e-foldings as

\begin{equation}
r= \frac{128\pi(1-f)}{1+f(N-1)}\,
\,\frac{\mu^2}{m_p^2}\left[1-\epsilon \left(\alpha+\beta A^2f^2
\left[\frac{Af}{1+f(N-1)}\right]^{\frac{2(1-f)}{f}}\right)^{-1/2}
\right]\,G_\gamma^2(N). \label{Rk11}\end{equation}

In Fig.(\ref{2}), the upper panel shows the evolution of the
scalar spectrum index $n_s$ versus the number of e-folds $N$, and
the lower panel shows the contour plot for the parameter $r$  as a
function of the $n_s$, for different values of the parameters $A$
and the ratio $m_5/\mu$. In particular, the dashed, dotted,  and
solid lines are for the pairs $(A=5.98\times10^{-2},m_5/\mu=0.2)$,
$(A=2.69\times10^{-2},m_5/\mu=0.02$, see Ref.\cite{PPP}), and the
standard intermediate model, respectively.  Here, we have used the
values $m_p=1$, $\eta=0.99$, $\epsilon=-1$, ${\cal{A}}_0=1$ and
$f=1/2$. From the upper panel, we observed that the number of
e-folds $N$ increases in the warped DGP scenarios when it is
compared with the the case of the standard intermediate model, in
which $n_s=1-(N+1)^{-1}$, see Ref.\cite{Barrow3}. From the lower
panel, we noted that for the $r-n_s$ graphs,  the ratio
$m_5/\mu=0.02$ presents a small displacement with respect to the
results obtained in the standard intermediate model
\cite{Barrow3}. Also,  we numerically found  that when we decrease
the value of the ratio $m_5/\mu$, for lower values, $m_5/\mu<0.02$
we observed that the new lines on the contour plot $r-n_s$ becomes
similar to the standard intermediate model.

\section{Conclusions \label{conclu}}

In this paper we have analyzed  the intermediate inflationary
model, for the cases of the brane model and the warped DGP model,
respectively. We have found solutions of the Friedmann equations
for a flat universe containing a standard scalar field $\phi$. For
both models, we considered the entire  Eqs.(\ref{HC}) and
(\ref{H2}), which show some  new attractive features, during
intermediate inflation. We have also obtained explicit expressions
for the corresponding, scalar field, effective potential, power
spectrum of the curvature perturbations $P_{\mathcal{R}}$,
tensor-scalar ratio $r$, and the scalar spectrum index $n_s$.
Also, in both cases, considering the WMAP seven year data, we have
found constraints on the parameters for our models. Here, we have
taken the constraint of the consistency relations, $n_s = n_s(r)$.

For the brane model, we noted that for values of the brane tension
$\tau<10^{-6}$, the model is well supported by the data as could
be seen from Fig.(\ref{1}). Here, we have used the values $m_p=1$,
$A\simeq 3.694\times 10^{-2}; 3.692\times
10^{-2};3.587\times10^{-2}$, and $f=1/2$, respectively. We
numerically observed that for the value of the tension of the
brane $\tau=10^{-6}$, the consistency relations, $n_s = n_s(r)$,
present a small displacement in relation to the standard
intermediate model. In this way, we noted that when we increase
the value of the parameter $\tau$, for values of the brane tension
$\tau>10^{-6}$, we observed that the contour plot in the $r-n_s$
plane becomes similar to the standard intermediate model.

For the warped DGP model, we noted numerically from
Eq.(\ref{nss1}) that the value of the spectral index is  $n_s\gg
1$ for the positive branch, $\epsilon=+1$, but this value of $n_s$
is disfavored from the observational data and the warped DGP model
does  not work for this branch. From the upper panel in
Fig.(\ref{2}), we observed that the number of e-folds $N$
increases in the warped DGP model when it is compared with the
standard intermediate model. From the lower panel we  numerically
noted that when we decrease the ratio $m_5/\mu$, for values
$m_5/\mu<0.02$, we observed that the new lines on the contour plot
$r-n_s$ becomes similar to the standard intermediate model. Here,
we have used the values $m_p=1$, $\eta=0.99$, $\epsilon=-1$,
${\cal{A}}_0=1$,$A\simeq 5.98\times 10^{-2}; 2.69\times 10^{-2}$,
and $f=1/2$.

We have also found in both models that the incorporation of the
additional term in  Friedmann's equation improves some of the
characteristics of the intermediate inflation. In particular, this
is emphasized in the consistency relations $n_s=n_s(r)$. Finally,
from the effective potentials obtained in both models (not present
minimum), we have not addressed the mechanism of reheating and
subsequent  connections to the standard Big-Bang model (see e.g.,
Ref.\cite{reheating}). Calculation for the reheating temperature
in theses scenarios would produce  new constraints on the brane
parameters. We hope to return to this point in the future.

\begin{acknowledgments}
R.H. was supported by COMISION NACIONAL DE CIENCIAS Y TECNOLOGIA
through FONDECYT grant  N$^0$ 1130628  and by DI-PUCV grant 123.724.
M.O. was supported by Proyecto D.I. PostDoctorado 2012 PUCV. N.V.
was supported by Proyecto Beca-Doctoral CONICYT N$^0$ 21100261.
\end{acknowledgments}


\end{document}